\documentclass{article}

\usepackage{arxiv}

\usepackage[utf8]{inputenc} % allow utf-8 input
\usepackage[T1]{fontenc}    % use 8-bit T1 fonts
\usepackage{hyperref}       % hyperlinks
\usepackage{url}            % simple URL typesetting
\usepackage{booktabs}       % professional-quality tables
\usepackage{amsfonts}       % blackboard math symbols
\usepackage{nicefrac}       % compact symbols for 1/2, etc.
\usepackage{microtype}      % microtypography
\usepackage{lipsum}		% Can be removed after putting your text content
\usepackage{graphicx}
\usepackage{natbib}
\usepackage{doi}
\usepackage{blindtext}
\usepackage{tcolorbox}
\usepackage{amsmath}
\usepackage{dutchcal} % mathcal support

\usepackage[toc,page]{appendix}

\usepackage{amssymb}
\usepackage{calligra,amssymb}

\title{Consideration Set Sampling to Analyze Undecided Respondents}

\author{\href{https://www.foundstat.statistik.uni-muenchen.de/personen/mitglieder/kreiss/index.html}{\includegraphics[scale=0.025]{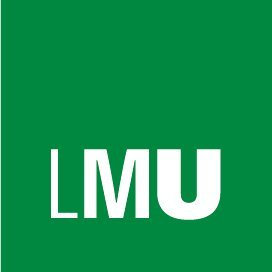}\hspace{1mm}Dominik Kreiss} \\
	Department of Statistics\\
	LMU Munich\\
	\texttt{dominik.kreiss@stat.uni-muenchen.de} \\
	\And
	\href{https://www.foundstat.statistik.uni-muenchen.de/personen/mitglieder/augustin/index.html}{\includegraphics[scale=0.025]{lmu.jpeg}\hspace{1mm}Thomas Augustin} \\
Department of Statistics\\
	LMU Munich\\
	\texttt{thomas.augustin@stat.uni-muenchen.de} \\
}

\renewcommand{\shorttitle}{Consideration Set Sampling to Analyze Undecided Respondents}

\hypersetup{
pdftitle={Consideration Set Sampling to Analyze Undecided Respondents},
pdfsubject={Statistics},
pdfauthor={Dominik Kreiss},
pdfkeywords={choice models, consideration set, pre-election poll, random set, undecided voters},
}

\newcommand\TAkom[1]{\relax}

\newcommand{\y}{\text{\calligra y}\,}
\newcommand{\Y}{\text{\calligra Y}\,}
\newcommand{\q}{\text{\calligra q}\,}
\newcommand{\s}{\text{\calligra s}\,}

\begin{document}
%\renewcommand{\autoref}[1]{\ref{#1}}

%%%%%%%%%%%%%%%%%%%zum Lesen  TA %%%%%
%\addtolength\textheight{-8cm}
%\addtolength\baselineskip{6ex}
%\addtolength\parskip{3ex}
%%%%%%%%%%%%%%%%%%%%%%%%%%%%%%%%%%%%

\maketitle
%%%%%%%%%%%%%%%%%%%%%%%%%%%%%%%%%%%%%%%%%%%%%%%%%%%%%%%%%%%%%%%%%%%%%%%%%%%%%%%%%%%%%%%%%%%%%
\begin{abstract}
% letzter Ansatz
Researchers in psychology characterize decision-making as a process of eliminating options. While statistical modelling typically focuses on the eventual choice, we analyze consideration sets describing, for each survey participant, all options between which the respondent is pondering. Using a German pre-election poll as a prototypical example, we give a proof of concept that consideration set sampling is easy to implement and provides the basis for an insightful structural analysis of the respondents' positions. The set-valued observations forming the consideration sets are naturally modelled as random sets, allowing to transfer regression modelling as well as appropriate machine learning procedures.

\end{abstract}
%%%%%%%%%%%%%%%%%%%%%%%%%%%%%%%%%%%%%%%%%%%%%%%%%%%%%%%%%%%%%%%%%%%%%%%%%%%%%%%%%%%%%%%%%%%%%
% keywords can be removed
 % \and Civey
%\TAkom{Grundsätzliche Änderungen\\ \\}
%\TAkom{Zitationsstyle geändert} 

%\TAkom{Zeilenumbrüche in Ttel geändert, oK?}
%%%%%%%%%%%%%%%%%%%%%%%%%%%%%%%%%%%%%%%%%%%%%%%%%%%%%%%%%%%%%%%%%%%%%%%%%%%%%%%%%%%%%%%%%%%%%

%%%%%%%%%%%%%%%%%%%%%%%%%%%%%%%%%%%%%%%%%%%%%%%%%%%%%%%%%%%%%%%%%%%%%%%%%%%%%%%%%%%%%%%%%%%%%
%%%%%%%%%%%%%%%%%%%%%%%%%%%%%%%%%%%%%%%%%%%%%%%%
\section{Introduction}
It is characteristic for human beings to ponder between options before finally making up their minds. At a given point in time prior to the eventual choice, individuals are often not yet determined but have already narrowed down their viable options to a compelling subset.
Rationals on how this set, often called \textit{consideration set}, emerges and which properties it satisfies have been developed in psychology and marketing research. (e.g.~\cite{StocchiConsideration,Shocker1991}) 
In this vain, one understands choice as a process of successively eliminating options by aspects, firstly excluding alternatives from the complete choice set and later on choosing from the emerged consideration set. \citep{tversky1972,Oscarsson2019}
Surprisingly, although substantial information concerning individual preferences and positions may be expected to lie in such consideration stages, most research has confined itself to the eventual choice.

A prototypical example of a choice process with a substantial amount of indecisiveness is voting in a multiparty system. Even though it is generally known that a relevant proportion of voters in multiparty systems is still undecided in pre-election polls (e.g.~\cite{Arcuri2008}), current polling designs do not allow those still undecided voters to provide their current position accurately. Indeed, these respondents are confronted with, respectively unsatisfactory, possibilities: they can give a spuriously precise answer by more or less arbitrarily selecting one of the parties from their consideration set; they can drop out; or they can tick the basically meaningless `don't know' category, comprising a mélange of the most diverse opinions. All these types of responses are detrimental, as either error is induced or valuable information is lost for a proper picture of the political landscape. 
As undecided voters appear to relevantly differ from the undecided as worked out by \cite{Oscarsson2019a}, simply neglecting them leads to a restricted and biased analysis.

%The underlying issue here is not restricted to elections but concerns other fields like consumer choice, such as narrowing down alternatives by brands or types of products.

In this paper, we argue that it is possible to exploit the valuable information in the consideration stage(s) in a comprehensive way, providing a vivid basis for insightful structural modelling of the respondents' preferences, positions, and attitudes. We advocate and introduce a direct {\em consideration set sampling}, where -- without the need to change the sample design itself -- the respondents are enabled to provide their current position by listing all alternatives they are still pondering between.\footnote{Asking participants for rankings of the options as suggested, for example, by \cite{furnkranz2010preference} does not address the underlying issue here, as the order would either induce a first choice or partial ranking leading again to set-valued data for the decision of interest.}
This extension has several appealing properties. 
\begin{itemize}
\item First of all, the information is collected in an undistorted way on an adequate level of coarseness, taking properly into account the fact that the respondents typically ponder between a few of the parties only and thus are neither completely indifferent nor already capable of stating one single party.
\item A second main advantage is the intuitive character of the answer for the undecided, who are, as argued above, naturally thinking in set-valued structures building on the choice process in stages (e.g.~\cite{Oscarsson2019a,Plass2015}) and thus directly find themselves with their position in the questionnaire. 
\item Thirdly, as concretely practised in this work in cooperation with the German polling institute Civey, implementing consideration set sampling in a survey proves to be relatively simple and cost-effective, as only one additional question is required to properly record the individual set-valued information.
\item And, last but not least, we show that it is immediately possible to embed methodologically the set-valued observations arising from consideration set sampling into the framework of so-called (conjunctive, onticly interpreted) random sets (e.g.,~\cite{Couso2014}). Since in this formalization process, the underlying random variable proves to be again of a categorical scale of measurement, popular methods for categorical data analysis like discrete choice modelling as well as standard machine learning procedures can successfully be transferred.  
\end{itemize}

Indeed, consideration set sampling provides a powerful basis for an insightful structural analysis of the respondents' genuine positions and attitudes, as we will demonstrate with the example of a German pre-election poll. Closely cooperating with one of the biggest German polling institutes Civey\footnote{More information is available under \url{https://civey.com/ueber-civey}, last visited 2023.07.25}, we are able to provide a carefully implemented first proof of concept of our methodological concepts. Eight weeks prior to the German 2021 Bundestag Election, 5076 voters participating in the survey have been asked for their consideration set, whereby 22.8\% indeed were still undecided, providing  a non-singleton consideration set. 
Our structural modelling of the political landscape based on the consideration sets, in particular, contains an adoption to our context of electoral choice models, several exemplary interpretable machine learning methods, and a cluster analysis as a prototypical unsupervised technique.  Thus, this work operationalizes, puts into practice, and statistically fertilizes fundamental theoretical insights into the choice process in general (e.g.~\cite{tversky1972}) as well in voting research (e.g.~\cite{Oscarsson2019}), hereby systematizing and substantially extending some first elementary corresponding deliberations presented at a symposium, a workshop and in a preliminary exploratory analysis prior to the election. \citep{Plass2015,Kreiss2020,Kreiss2021b}

Concretely, this paper proceeds as follows:
In Section~\ref{ontic}, we formalize consideration set sampling by random sets and show under which circumstances methodology can generally be adapted for structural analysis. 
Then, in Section~\ref{modeling}, we suggest in our framework some explicit approaches. This includes regression-based modelling (Section~\ref{theory:regression}) as well as techniques from interpretable machine learning (Section~\ref{theory:iml}) and unsupervised learning (Section~\ref{theory:unsupervised}).  Afterwards, in Section~\ref{application}, we apply the discussed methods to real data collected in cooperation with Civey and show how new insights can be obtained.
In the concluding remarks in Section~\ref{ch_concluding_remarks}, we reflect on further avenues and opportunities enabled by consideration set sampling.

%\include{Tex/old_ch1} % hier noch die alte version 

%%%%%%%%%%%%%%%%%%%%%%%%%%%%%%%%%%%%%%%%%%%%%%%%%%%%%%%%%%%%%%%%%%%%%%%%%%%%%%
\section{Methodical Framework for a Structural Analyses of Consideration Sets}\label{sec:methods}
%%%%%%%%%%%%%%%%%%%%%%%%%%%%%%%%%%%%%%%%%%%%%%%%%%%%%%%%%%%%%%%%%%%%%%%%%%%%%%

%%%%%%%%%%%%%%%%%%%%%%%%%%%%%%%%%%%%%%%%%%%%%%%%%%%%%%%%%%%%%%%%%%%%%%%%%%%%%%
\subsection{Consideration Set Sampling and Random Sets}\label{ontic} 

To describe and implement our consideration set sampling, consider a finite set ${\mathcal S}=\{a, b, c, \ldots\}$ of unordered elements, for instance, the major parties in a pre-election poll in a multi-party system, $P({\mathcal S})$ its power set, and a space ${\mathcal X}$ collecting potential values of the covariates, in particular, socio-demographic characteristics in our example. We sample $n$ units from the underlying population $\Omega$, whereby we suppose the population to be large enough to allow for treating it as infinite, neglecting specific finite population effects. The deliberations in the sequel rely on simple random sampling (with replacement) producing i.i.d.~random variables/random elements.\footnote{Techniques to adopt to complex sampling designs (e.g., \cite{skinner_introduction_2017}) should be transferable in principle; a detailed discussion of this issue, however, is left to further research.} 
For technical reasons or context-dependent choices, the set of all options of the power set usually has to be reduced to a subset, denoted as $\tilde{P}({\mathcal S })$, such that $\tilde{P}({\mathcal S }) \subset P({\mathcal S })\setminus\emptyset$ and $\{q\}\in \tilde{P}({\mathcal S }),$ for all $q\in \mathcal{S}$. % excluding the empty set. Throughout the paper, we assume $P(S)$ to contain all singletons. % (ggf noch hoch und einführen Begriff reduced power set)
%This leads to reamining $|\tilde{{\mathcal P}}({\mathcal S })| = k$ options.

We hence obtain realizations $(\y_1,x_1),$ $\ldots,$ $(\y_i,x_i),$ $\ldots,$ $(\y_n,x_n)$ of the random elements\footnote{Measureability is not explicitly problematized here. Working with an underlying measurable space $(\Omega, {\mathcal A})$ and an appropriate $\sigma-$field ${\mathcal F}$ over ${\mathcal X}$, we rely canonically on the smallest $\sigma-$field generated by $P(\tilde P({\mathcal S})) \times {\mathcal F}$.}
\begin{eqnarray}
(\Y_i, X_i) \,:\,  \Omega & \longrightarrow & \tilde{P}({\mathcal S }) \times {\mathcal X} \\
 &&\qquad \qquad \qquad \qquad  i=1,\ldots,n,\label{230709-variables} \nonumber \\[-3ex]
\omega_i &\mapsto & (\y_i,x_i)\,.
\end{eqnarray}
The random elements $\Y_i$ are random sets (e.g., \cite{Molchanov2005}), formalizing the crucial aspect that the observed considerations sets typically are set-valued, i.e.~consisting of several elements of ${\mathcal S}$ and thus being an element of the reduced power set $\tilde{P}({\mathcal S })$. For the sake of a unified representation, the response of  an already decided respondent is described as a singleton observation, i.e.~the already taken decision for party $q\in\mathcal{S}$ is identified with the consideration set $\{q\}$. % $ \in \tilde{P}(\mathcal{S})$. 

Random sets have two fundamentally different interpretations. Following the terminology in \cite{Couso2014}, an {\em ontic} and an {\em epistemic} view have to be distinguished. The ontic view understands a set-valued observation as a holistic entity and thus as an irreducible, precise observation per se. The epistemic view, in contrast, sees the set as an imprecise, coarsened description of a genuinely precise outcome, which would correspond to the eventual choice in our context. Apart from some brief remarks in the Concluding Remarks 
in Section~\ref{ch_concluding_remarks}, we focus throughout our paper on the ontic point of view, being interested in a structural analysis of indecision between certain parties as a specific political position of its own. Consequently, set relations between the consideration sets are (taken as) meaningless: For instance, indecision between two parties $a_1$ and $a_2$  is in no way set-theoretically related to indecision between $a_1$, $a_2$ and $a_3$; the consideration sets $\{a_1,a_2\}$ and $\{a_1,a_2,a_3\}$ reflect two different, mutually unrelated positions of their own. Technically, this means that -- with the underlying space of options ${\mathcal S}$ being categorial -- $\tilde{P}({\mathcal S})$ is categorical as well, making the random sets $\Y_i$, $i=1,\ldots, n,$  categorical random elements, to which statistical methods for analyzing categorical variables, therefore, can be transferred. We make direct use of this crucial methodological fact. Thus, in the sequel we build our structural analysis on categorical regression with outcome space $\tilde{P}({\mathcal S})$ as well as on interpretable learning methods with $\tilde{P}({\mathcal S})$ as their target space and unsupervised learning techniques with underlying space $\tilde{P}({\mathcal S }) \times {\mathcal X}$. For the sake of clarity, we will rely on the following notational convention: small standard Latin letters denote elements of ${\mathcal S}$ (cf.~above), while small calligraphic letters like $\q$ stand for elements of $\tilde{P}({\mathcal S})$. %%%%%; i.e.~typical events have the form $\{Y=q\}$ and $\{\Y=\q\} $. %%%TA wir haben praktisch kein Y mehr
%(l ändern in q? erkennt man genausowenig, evtl da such den anderen stiel? gar nicht so einfach.)

%%%\input{Tex/2.1_old}

%%%%%%%%%%%%%%%%%%%%%%%%%%%%%%%%%%%%%%%%%%%%%%%%%%%%%%%%%%%%%%%%%%%%%%%%%%%%%%
%%%%%%%%%%%%%%%%%%%%%%%%%%%%%%%%%%%%%%%%%%%%%%%%%%%%%%%%%%%%%%%%%%%%%%%%%%%%%%
%%%%%%%%%%%%%%%%%%%%%%%%%%%%%%%%%%%%%%%%%%%%%%%%%%%%%%%%%%%%%%%%%%%%%%%%%%%%%%
\subsection{Structural Analysis based on Consideration Set Sampling}\label{modeling}
%%%%%%%%%%%%%%%%%%%%%%%%%%%%%%%%%%%%%
% Überblick
Using the set-valued information about the current position, we want to gain a more comprehensive picture in our example of the political landscape and political positions. Building on the general concepts developed in the previous section, we now provide explicit modelling approaches demonstrating how applied research can benefit from consideration set sampling. 
We first discuss the approaches in a general setting and then apply them to our data set.
Hereby we suggest methods from regression-based choice modelling as well as interpretable and unsupervised learning, all based on the 
%As a framework, we suggest methodology from regression-based ones, the field of interpretable machine learning, and unsupervised or content-related models. 
i.i.d.~sample $(\y_1,x_1),$ $\ldots,$ $(\y_i,x_i),$ $\ldots,$ $(\y_n,x_n)$.%%%TA mehr würde ich hier nicht mehr sagen; der Rest ist oben integriert.
%from $(\Y_i, X_i)$, where the restriction $\tilde{P}({\mathcal S })$ is fixed for each analysis. 
%The random element $\Y_i \in \tilde{P}({\mathcal S })$ hereby has a number of $|\tilde{P}({\mathcal S })| = : k$ possible outcomes. % denoted with $\mathcal{l}$. (An TA: ggf ganz ohne k? Oder doch?) 

%We provide a methodological framework for how to approach this unique situation with the arising state space while the models can be modified in a context-dependent manner.  

%%%%%%%%%%%%%%%%%%%%%%%%%%%%%%%%%%
\subsubsection{Multinomial Regression Approaches}\label{theory:regression}

As argued in Section~\ref{ontic}, the random sets $\Y_i$, $i=1,\ldots,n,$ are categorical random elements, and thus we can directly transfer the common multinomial settings to our situation. We will briefly investigate the marginal distribution of the considerations sets, i.e.~the positions $\Y_i$, and then turn to regression-based approaches, where the positions are seen as dependent and the covariates $X_i$ as independent variables.

We firstly look at the marginal distribution of $\Y_1,\ldots, \Y_n$ with $\pi_\q\stackrel{{\rm def}}{=}P(\{\Y_i=\q\})$, $\q \in \tilde{P}(\mathcal{S})$, and $\sum_{\q \in \tilde{P}(\mathcal{S})} \pi_\q =~1$. The underlying samples can be summarised by an appropriate count statistic $(N_{\q})_{\q\in\tilde{P}(\mathcal{S})}$ with $N_{\q} \stackrel{{\rm def}}{=} |\{i | \Y_i = \q\}|$ reflecting how many respondents state category $\q \in \tilde{P}(\mathcal{S})$. With the corresponding sample denoted by $(n_{\q})_{\q\in\tilde{P}(\mathcal{S})}$, this results in a multinomial likelihood $$ {\rm lik}\big((\pi_\q)_{\q \in \tilde{P}(\mathcal{S})} \big|\big| (n_{\q})_{\q\in\tilde{P}(\mathcal{S})}\big) \propto\prod_{\q \in \tilde{P}(\mathcal{S})} (\pi_\q)^{n_{\q}}\,,$$ yielding naturally the relative frequencies $\frac{n_{\q}}{n}$ of the observed positions $\q$ as maximum likelihood extimates of the corresponding probabilities $\pi_\q$. 
%\mathbf{N}=, \mathbf{n}=

%We can then construct the likelihood over those counting variables in the sense: 
%
%\begin{equation}
%    f(\mathcal{n}_1, \cdots, \mathcal{n}_{|\tilde{P}({\mathcal S })|}) \propto  \pi_1^{\mathcal{n}_1} \dots  \pi_{|\tilde{P}%({\mathcal S })|}^{\mathcal{n}_{|\tilde{P}({\mathcal S })|}} 
%\end{equation}
%\noindent
%with $\pi_{\q} := P(\Y_i = \q)$ and the restriction $\sum^{|\tilde{P}({\mathcal S })|}_{l=1} \pi_l = 1$, which enforces using a %reference category or asymmetric constraint for the $\pi_l$ to be probabilities.

We base our regression analysis on the adaption of multinomial regression models, which are common in marketing and voting research  (e.g.~\cite{TutzBook,HensherDCM}) as they enable a natural linear interpretation of the coefficients. For our categorical random elements, we can then write the multinomial logit model following \cite[p.~211 ff.]{TutzBook} with linear predictor consisting of the covariate vector $x_i$ together with the category-specific parameters $(\beta_\q)_{\q\in \tilde{P}({\mathcal S }) }$ in its generic form as:
\begin{equation}
    P(\Y_i = \q| x_i) = \frac{\text{exp}(x_i^T \beta_{\q})}{\sum_{\s\in \tilde{P}({\mathcal S } )}  %^{|\tilde{P}({\mathcal S })|}_{l = 1} 
    \text{exp}(x_i^T \beta_\s)}\,,
\end{equation}
whereby fixing a reference category or symmetric side constraint is necessary to ensure well-definiteness.  Estimation can be obtained via maximum likelihood or along the Bayesian paradigm.%, whereby methods can be further combined with regularization. 
%%%Satz nach unten verschoben TA

In most settings with consideration set sampling, the number of potential groups is rather large,   and even after the original reduction from $P(\mathcal{S)}$ to $\tilde{P}(\mathcal{S)}$ observations might get stretched thin.
From an applied standpoint, it is recommended to limit the degrees of freedom needed for example by additional regularization.
%Hence, regularization is advisable as well as it can be better to further reduce groups context-dependent.

% Regularization 
Such regularization can be conducted with different concepts. 
With the penalization parameter $\lambda$, the penalized log-likelihood $pl(\beta)$ can be written in the general form with regards to the usual log-likelihood contribution of the \mbox{$i-$th} observation ${\rm l}_i(\beta)$ as: ${\rm pl}(\beta) = \sum^n_{i = 1} l_i(\beta) - \frac{\lambda}{2}J(\beta)$ with $J(\beta)$ as a function penalizing the parameters. \cite[p.~233 ff.]{TutzBook}
Both $L1$ and $L2$ penalization are possible as well as it can be beneficial to group coefficients to remove some covariates entirely, following \cite{Tutz2015} or \cite{vincent2014sparse}. 
% Besprechen TA: Eine Regularisierung aufschreiben, oder das im Anwendungsteil
The penalization parameter $\lambda$ is usually determined by cross-validation, but relying on the Akaike or Bayesian information criterion is also possible. 
% P Wert
All covariates outlasting the regularization do have some contribution but are not necessarily significantly different from zero. Confidence intervals and hypothesis testing are impeded by regularization but not impossible. (see e.g.~\cite{inference_on_regularized_regression})

% General Extesnsions
%%%%  %TA gelöscht da oben schon genauso gesagt %%% Depending on the covariates' structure, extensions to additive models and explicitly including ordinal covariates in the estimation process are possible. Approaches with alternative specific coefficients in Discrete Choice Models can be included if available. 
% Final sentence
%von oben geholt 
An adequate model has to be fitted context-dependent. Further extensions to additive models 
(e.g.~\cite{ravikumar2009sparse}), ordinal covariates or alternative specific coefficients  
(e.g.~\cite{HensherDCM}) can be incorporated as well. With such a regression model, characteristics of the interesting positions can be analyzed, providing a new opportunity to gain empirically founded insights about the undecided.

%%%%%%%%%%%%%%%%%%%%%%%%%%%%%%%%%%%%%%%%%%%%%%
\subsubsection{Interpretable Machine Learning Methods}\label{theory:iml}

The field of interpretable machine learning provides a range of approaches to examine structural properties in this data situation parallel to the regression approaches. % parallel to the regression ones.
For this, we learn a classifier amongst the options of $\Y_i$ with observations $(\y_1,x_1),$ $\ldots,$ $(\y_n,x_n)$, in the classical supervised learning setting and then examine according to which structures the model classifies. 
The advantage of supervised machine learning lies in the relaxation of model classes, which allows for non-linear relations and (higher-order) interactions between the covariates. Additionally, ordinal covariates and restrictions with degrees of freedom can be easier included.
A carefully chosen learner should hence represent the connection between the positions and covariates more flexibly than the regression approaches suggested above, however, at the cost of the straightforward interpretation. 
In order to regain (some of) the interpretability, several approaches have been developed. 
(e.g.~\cite{molnar2020interpretable})
They cover global methods like partial importance (PI) or the permutation feature importance (PFI) and local methods like \textit{Local Interpretable Model-Agnostic Explanations} (LIME) or \textit{SHapley Additive exPlanations} (SHAP). \citep[p.~2]{molnar2020general}
Depending on the underlying question, different approaches are worth considering.

% UNser ansatz und das algemeiner   
%In our case, the underlying model can be trained for classification amongst all groups or only comparing fewer or two positions of particular interest. 
%Different subgroups from the space $\mathcal{P(S)}$ can be analyzed, in particular, only binary ones to compare two groups. 
For the corresponding model, it is then possible to look at different approaches and model classes.
While the Feature Importance gives an immediate intuition about the influence of covariates and the Partial Dependence Plots provide a nice but simplified picture, the local and more complex approaches are more thorough.   
For example, \textit{SHAP-Values} introduced by \citep{lundberg2017} from coalitional game theory, try to determine the contribution of one feature to the overall prediction. 
%With a kernel trick, these (otherwise reliant on vast computations) Shap-Values become computationally feasible. % ggf noch Lime 
% Quality aspects
One has to keep in mind that the results are reliant on the applied model and, with it, inevitably on the underlying data. 
Fitting problems, bad generalizations, or issues with imbalanced data must be addressed, due to their influence on the interpretation. \citep[p.~5]{molnar2020general}
% Imbalanced Data 
Especially, as with increasing groups, the data becomes more imbalanced one has to be cautious with statements about the smaller groups, and oversampling methods like \textit{SMOTE} \citep{Chawla2002} can be considered in the multi-group case.  
% Our goal
If the learner is chosen and tuned correctly, these approaches provide an interesting and additional insight into patterns and connections between the political positions and the covariates of interest.

%%%%%%%%%%%%%%%%%%%%%%%%%
\subsubsection{Unsupervised Learning and Partitions of the Population}\label{theory:unsupervised}

% Unsupervised 
Unsupervised machine learning approaches take a further and different look at the data situation of the underlying space $\tilde{P}\mathcal{(S) \times X}$. 
With this, we are interested in finding patterns and structural peculiarities in the data with a connection to the different groups of undecided participants. 
One interesting way to illustrate the partitioning of the underlying population is first to determine socio-demographic clusters and then examine the choice positions in these clusters. 
This gives an impression and intuition about the interconnection of the positions and the covariates.
There are several ways to determine clusters in the population. Either they are already given due to content-relate restrictions, or machine learning approaches can be applied for the determination. (e.g.~\cite{ghahramani2003unsupervised,hastie2009unsupervised})
This is one way to discriminate the participants according to their features and illustrates to what extent distinctive groups can be associated with particular political positions.
Another idea is to regard particular natural population partitions from sociological or politological research and then illustrate the groups amongst these strata.  
\\\\
%%%%%%%%%%%%%%%%%%%%%%%%%%%%%%%%%%
% Outlook
In the following chapter, we implement approaches from all three methodological categories to a pre-election poll for the 2021 German federal election, demonstrating how new insights can be obtained due to the accurate representation of undecided voters. 

%\input{Tex/epistemic}

%%%%%%%%%%%%%%%%%%%%%%%%%%%%%%%%%%%%%%%%%%%%%%%%%%%%%%%%%%%%%%%%%%%%%%%%%%%%%%
\section{Application to the 2021 German Federal Election}\label{application}
%%%%%%%%%%%%%%%%%%%%%%%%%%%%%%%%%%%%%%%%%%%%%%%%%%%%%%%%%%%%%%%%%%%%%%%%%%%%%%
%%%%%%%%%%%%%%%%%%%%%%%%%%%%%%%%%%%%%%%%%%%%%%%%%%%%%%%%%%%%%%%%%%%%%%%%%%%%%%
\subsection{The Data}
We apply exemplary approaches from all three methodological categories discussed above to the sample two months before the 2021 German federal election. 
A two-stage survey format with consideration set sampling was developed in cooperation with the polling institute Civey\footnote{See also the news article about our cooperation under \url{https://civey.com/ueber-civey/unsere-methode/artikel/civey-unterstuetzt-forschung-zur-mitberuecksichtigung-unentschlossener}, last visited 23.07.23}, collecting the set-valued information for the undecided next to 11 categorical or ordinal features about socioeconomic status and demographics.
% Two stages
First, participants are distinguished whether or not they are uncertain about their vote and, in a subsequent step, asked about either that particular party or all of their viable options. 
Unfortunately, no questions concerning positions on political issues were included in the survey, despite their perceived increasing importance for voters' choice. %The connection between demographics and socioeconomic status with the vote can still be observed in our dataset. %, but not with the same magnitudes as some decades ago. (e.g. \cite{Dippel2022})
\\
% Background on Covariates and Parties
In Germany, there are currently six relevant parties likely to surpass the 5\% hurdle (typically) necessary for being eligible for a seat in parliament.\footnote{For more information about the German voting system see: \url{https://www.bundeswahlleiterin.de/bundestagswahlen/2021/informationen-waehler/wahlsystem.html}, last visited 2023.07.25} 
Those are: \texttt{The Left, Green Party, SPD, CDU/CSU, FDP, AfD}.\footnote{Note that CDU/CSU is seen as one party here, even though it is assembled from different regions of Germany}
There is no natural ordering of the parties as classification like left-to-right scale or liberal-conservative somewhat lost meaning in the shifting political spectrum. (e.g.~\cite{Dippel2022})
We focus only on the main parties here, resulting in a six-dimensional initial space $\mathcal S$.

% Representative and Weighting
Civey constructs online surveys, with the benefits of large sample sizes but the downside of an initially non-random sample of the voting population.
To address this, they implement a post-stratification process by sub-sampling from an initial five times bigger sample to achieve approximate representativity.\footnote{Further information about the Civey procedure can be found in \cite{Richter2022}}
Regardless of the potential error induced hereby and possibly missing not at random structures, Civey established itself with this procedure amongst other polling institutes in Germany. They hence provide us with a state-of-the-art sample with 5076 observations two months before the German federal election, which we treat as an i.i.d.~sample in coherence with Civey's post-stratification in the following applications. 

%%%%%%%%%%%%%%%%%%%%%%%%%%%%%%%%%%%%%%%%%%%%%%%%%%%%%%%%%%%%%%%%%%%%%%%%%%%%%%
\subsection{Application}\label{application_data}

In this section, we illustrate how the modelling approaches in chapter \ref{modeling} provide interesting insights into both decided and undecided voters two months prior to the German federal election.
While the basic methodological considerations are our main interest, we also provide brief interpretations of the results for each corresponding approach without pushing us in the field of political research.
% ggf variances: In our data, the socioeconomic variables tend to have a rather small connection to the political position.  

%%%%%%%%%%%%%%%%%%%%%%%% 
\subsubsection{Applied Grouped Regularized Regression}\label{methods:regression}

Starting with the regression approach, we require a feasible minimum of observations in each group as well as sparse covariates to avoid perfect separation. 
We examine the six relevant parties together with the three biggest consideration sets of the individuals undecided between \texttt{SPD/Green/Left}, \texttt{SPD/Green} and \texttt{CDU/CSU/FDP} for an overview. 
%Already those nine groups, together with the categorical covariate structure leads to high numbers of parameters to be etimated. 
%In our data, we have six major parties and three consideration sets of particular interest with more than 50 observations each. Together with the categorical and ordinal covariates leading to 32 parameters for each group, the full model would contain overall 288 parameters to estimate. 
%Looking at all groups together as we do in this exemplary approach has a nice overview character and is possible with multinomial models, but a more focused analysis of two particular groups of interest is viable as well.  
For this, we binarize covariates and employ grouped L1 regularization from the \textit{glmnet} package \citep{glmnet} and determine the regularization parameter $\lambda$ with cross-validation. 
This reduces the number of the coefficients illustrated in 
Table~\ref{fig_msgl}, which can be interpreted as log odds with a symmetric constraint to enable estimation.\footnote{As mentioned above, hypothesis testing is somewhat controversial in combination with regularization. We did not provide confidence intervals.} 
Due to the grouping, four covariates were regularized precisely to zero.

\begin{table}[ht]
\resizebox{\textwidth}{!}{%
\centering
\begin{tabular}{rrrrrrrrrr}
  \hline
% & Left & \begin{tabular}[c]{@{}ll@{}}SPD/\\Green/\\Left\end{tabular} & Green & \begin{tabular}%[c]{@{}l@{}}SPD/\\Green\end{tabular}& SPD & \begin{tabular}[c]{@{}l@{}}CDU/\\CSU\end{tabular} & %\begin{tabular}[c]{@{}ll@{}}CDU/\\CSU/\\FDP\end{tabular} & FDP & AfD \\ 
& Left & \begin{tabular}[c]{r}\ SPD/\\Green/\\\ Left\end{tabular} & Green & \begin{tabular}[c]{r}SPD/\\Green\end{tabular}& SPD & \begin{tabular}[c]{r}CDU/\\CSU\end{tabular} & \begin{tabular}[c]{r}\ CDU/\\\ CSU/\\\ FDP\end{tabular} & FDP & AfD \\ 
 \hline
(Intercept) & -0.30 & -1.47\ \  & 0.81 & -0.93\ \ & 0.63 & 1.58 & -1.11\ \ \ \  & 0.05 & 0.73 \\ 
  City Habitant & 0.01 & 0.03\ \ & 0.09 & -0.02\ \  & 0.08 & -0.07 & -0.03\ \ \ \  & -0.00 & -0.08 \\ 
  Low Purchasing Power & 0.04 & 0.01\ \  & -0.04 & -0.01\ \ & 0.03 & -0.06 & -0.02\ \ \ \  & 0.00 & 0.04 \\ 
  University Degree & 0.02 & 0.01\ \  & 0.18 & -0.00\ \  & -0.09 & -0.06 & 0.04\ \ \  \ & 0.03 & -0.13 \\ 
  Married & -0.09 & -0.11\ \  & -0.12 & 0.01\ \  & -0.01 & 0.20 & 0.03\ \ \ \  & 0.04 & 0.05 \\ 
  Non-Religious & 0.19 & 0.01\ \  & 0.07 & -0.02\ \  & -0.10 & -0.32 & -0.03\ \ \ \  & 0.02 & 0.19 \\ 
  Former West Germany & -0.44 & 0.09\ \  & 0.37 & 0.16\ \  & 0.30 & -0.06 & 0.10\ \ \ \  & 0.05 & -0.56 \\ 
   \hline
\end{tabular}}
    \caption{Results from a (group)-regularized multinomial logit model with symmetric constraint and binarized covariates. Four covariates are regularized precisely to zero and are not included in the table.}
    \label{fig_msgl}
\end{table}

%\begin{figure}[ht!]
%    \centering
%    \includegraphics[width = \textwidth]{Tex/Plots/results_glmnet}
%\end{figure}

% Interpretation    
Two major insights occur, stressing the benefits of including the undecided voters. 
First, we are now capable of analyzing the groups undecided between specific parties. % gaining valuable new insights into the political landscape.
Second, we can determine differences between the positions of the single parties and consideration sets containing those parties. This shows how individuals undecided between specific parties really constitute new groups.
If we, for example, take the coefficient of the covariate \texttt{Married}, we see a slightly positive coefficient for being undecided between the \texttt{SPD} and the \texttt{Green Party}, while the one for those determined to vote for \texttt{Green} and those undecided between the \texttt{SPD}, \texttt{Green}, and \texttt{Left} is negative. The \texttt{Univerity Degree} seems to distinguish the \texttt{Green Party} from those undecided between the \texttt{SPD} and \texttt{Green Party}.
By examining various variables, both differences and similarities can be identified. This shows that including groups undecided between specific parties contributes something new, highlighting interesting aspects hidden before.

%%%%%%%%%%%%%%%
\subsubsection{Applied Interpretable Machine Learning}\label{methods:IML}

We now focus on two groups of particular interest and exemplary employ SHAP-values for structural insights here. 
We take the perspective of the \texttt{Green Party} and follow a strategically important question: What distinguishes its convinced supporters from those who are also considering voting for my biggest rival, the \texttt{SPD}?

As the underlying model, we choose a gradient boosting tree for binary classification between \texttt{Green Party} and \texttt{SPD/Green}, implemented with the package \texttt{h2o} \cite{h2o_R_booklet}, leading to an error rate of 4.4\% on the training and 9.6\% on the test data in our exemplary setting. 
We subsequently investigate how the model made the classification choice by looking at the SHAP-values. 

%The groups are rather imbalanced, with 106 to 793 observations, and the classifier has 
% SHAP VALUES
\begin{figure}[ht!]
\centering
	\includegraphics[width = 0.8\textwidth]{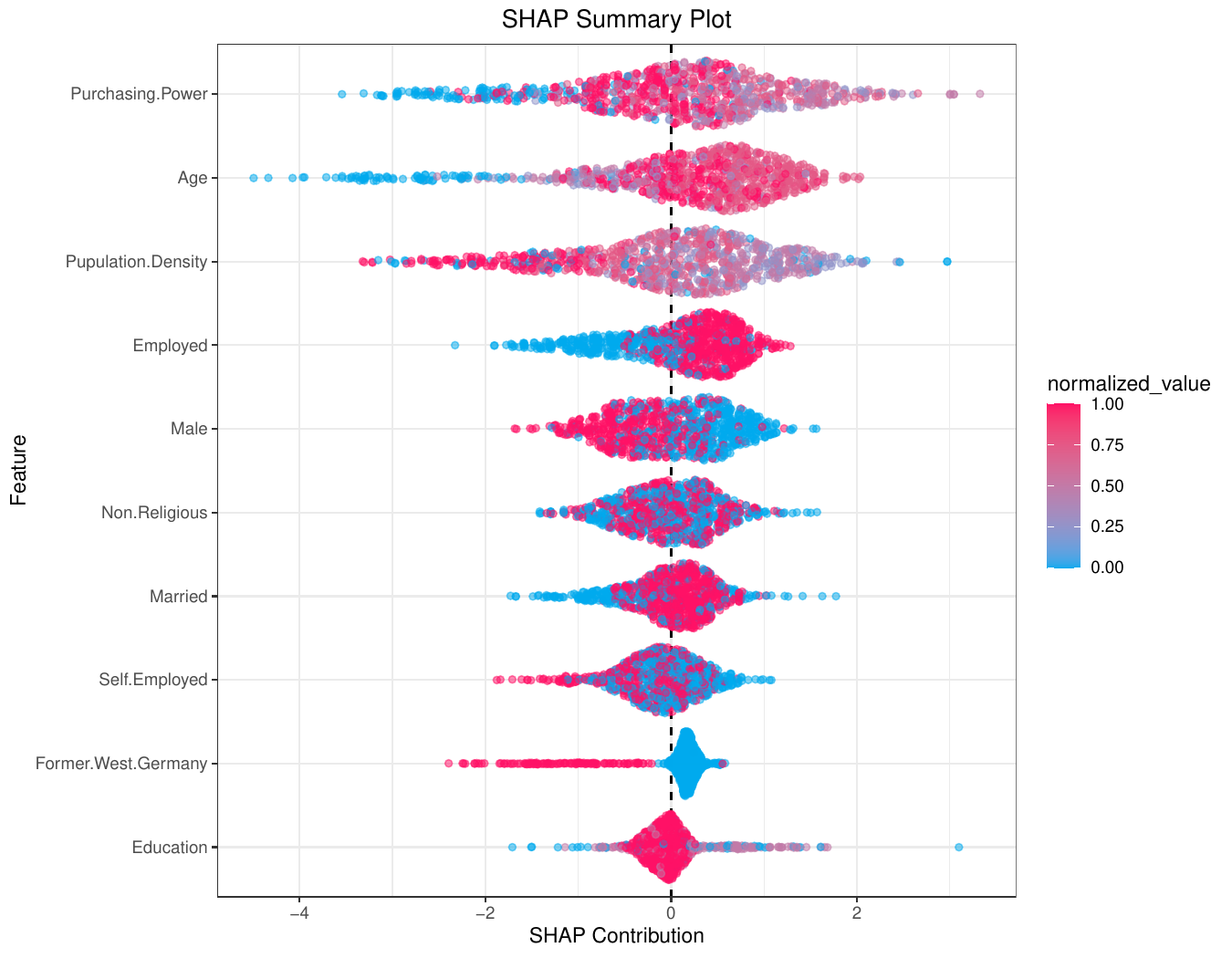}
	\caption{SHAP values for all covariates for every observation in the data. For each covariate, the associated SHAP-value is plotted on the x-axis. Each point represents a single model decision. The more influential a feature is, the more negative or positive its associated SHAP-value. A red point goes in the direction of being undecided between the \texttt{SPD/Green}}\label{shap}
\end{figure}
A summary plot of the SHAP-values is given in Figure~\ref{shap}. Here we see the SHAP-based connection of the corresponding variable to the binary classification. The extent of the relation is visualized but without communicating the attached uncertainty. 
% Overall conclusion 
According to the model, the covariate \texttt{purchasing power} has the strongest connection to the model's decision. Especially those with a medium \texttt{purchasing power} seem to be more often classified as undecided. 
Furthermore, a good distinction can be observed with the covariate \texttt{male}.
Results like this, potentially accompanied by thorough descriptive analysis, provide essential information for election campaign strategy and give potential starting points to orientate a campaign at. They are furthermore of interest to the neutral observer as well.

%%%%%%%%%%%%%%%
\subsubsection{Applied Unsupervised Lerning}\label{methods:unsupervised}

\begin{figure}[ht!]
\centering
	\includegraphics[width = 0.5\textwidth]{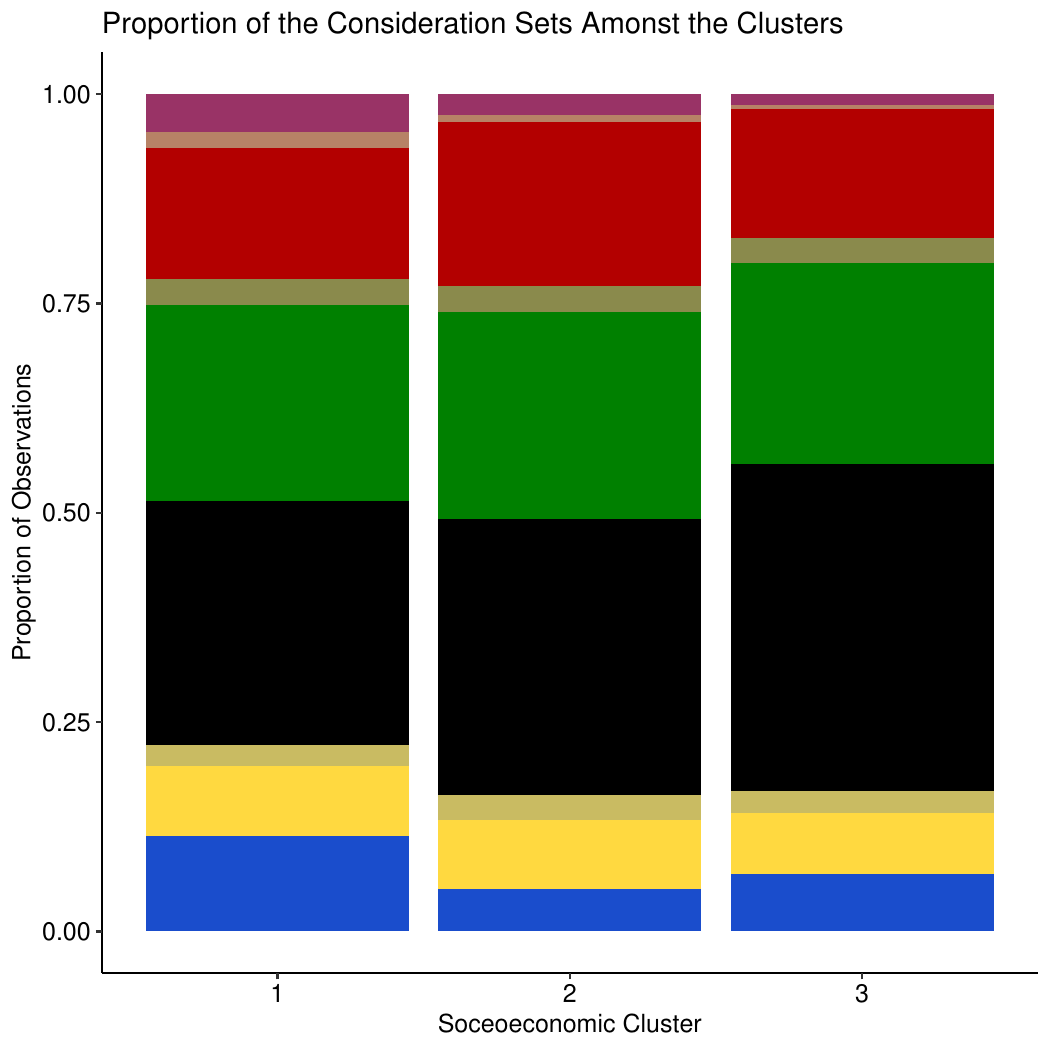}\\
	\includegraphics[width = 0.4\textwidth]{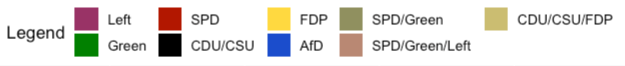}
	\caption{Three socio-demographic clusters determined by tree-based dissimilarity. In each cluster, the proportion of decided and undecided survey participants is shown.}\label{cluster_rf}
\end{figure}

We now take a broader look at different socioeconomic classes in our dataset. 
We strive for a simple overview of the political positions' connection to socioeconomic classes.
To this end, we employ an unsupervised machine learning model relying on tree-based dissimilarity following \cite{shi2006unsupervised}, to establish three different groups with the 11 covariates available as discussed in Section~\ref{theory:unsupervised}.
These three groups are now examined concerning the proportions of political positions. 
We illustrate the results in Figure \ref{cluster_rf}, with the proportions of political groups in each cluster.

\noindent
We get intuition and first impressions about the covariates and their connections with the political positions. 
There are some differences between the groups like a slight overrepresentation of individuals favoring the \texttt{Left Party} in the first cluster, mostly consisting of younger and well-educated participants, and a diminishing proportion of individuals indifferent between the \texttt{SPD}, \texttt{Green} and \texttt{Left} in the third cluster. % nochmal in die Daten schauen und auch set.seed
Overall some differences occur but the clusters seem rather similar, which hints towards a relatively weak discriminative power of the features.  
This supports the idea currently discussed in public that socioeconomic status and demographics do not determine political positions as strongly as it has some decades ago in Germany.
The clusters are quite intermixed, hinting that other properties besides socioeconomic ones determine the election choice. One could guess that the polarization within society induced by the pandemic as well as the migration crises of 2015 in Germany, plays a key role here and should be examined in further research. 
Overall the data gives the impression that socioeconomic status does indeed play a role but only explains the choice process of voting to a rather limited extent.

%%%%%%%%%%%%%%%%%%%%%%%%%%%%%%%%%%%%%%%%%%%%%%%%%%%%%%%%%%%%%%%%%%%%%%%%%%%%%%
\section{Concluding Remarks}\label{ch_concluding_remarks}
%%%%%%%%%%%%%%%%%%%%%%%%%%%%%%%%%%%%%%%%%%%%%%%%%%%%%%%%%%%%%%%%%%%%%%%%%%%%%%

% What we did
Motivated by the common neglect of undecided survey participants, we developed the framework of consideration set sampling for comprehensive structural analysis.
By allowing a set-valued characterization of the relevant group of not yet fully decided respondents, we obtained an accurate representation of their current position that can be interpreted as conjunctive random sets. 
With this, the established methodology can be successfully transferred. % to the new state space satisfying the same mathematical properties from our modeling point of view.
%This simple yet pivotal idea allows for a broader picture of the population. 
We developed a collection of methodological suggestions tailored to the new situation containing regularized regression and interpretable and unsupervised machine learning.  
In our application to pre-election polls with a self-constructed survey, a more complete picture of the political landscape arises. 

% Further work
This work can be extended methodologically in multiple directions. 
% Epistemic
Most evidently, approaches under the epistemic view, utilizing consideration sets as  partial information to improve forecasting and predictions about the eventual choice, can be developed. First ideas were introduced by \cite{Plass2015,Kreiss2020,Kreiss2020b}, but, taking a broader perspective, the framework of consideration set sampling can be embedded both in the theory of learning from imperfect data in statistics and machine learning. 
In the machine learning context, one could follow up ideas on \textit{Partial Label Learning} \citep{partial_label_learning}, \textit{Multi-Label Learning} \citep{Zhang2014}, or \textit{Superset Learning} \citep{Huellermeier2014} as explored in \cite{rodemann2022levelwise}.
In statistics, the limitations of the strong assumption of  \textit{Coarsening at Random} \citep{Heitjan1991} are interesting. 
In this setting, a tradeoff between the plausibility of the underlying assumptions and the conciseness of the results has to be addressed, following Manski's law of decreasing credibility \cite[p.~1]{Manski2003}, opening up avenues for approaches with partial identification \citep[e.g.~][]{MOLINARI2020355,arpino2014using,Li:Millimet:Roychowdhury:2023}.

% Longitudinal
As a connection between the ontic and the epistemic view, analyzing the shifts of consideration sets over time in a longitudinal manner could provide further new insights into the choice process. 
%Consideration set sampling over different points in time could als

% Outlook
We believe that interpreting consideration set as conjunctive random sets gives undecided respondents appropriate statistical representation. The advantages of structural analysis are evident: simple implementation, undistorted information collection as well as new insights on socio-demographic determinants of both undecided and decided voters.
%compared to the common neglection while utilizing this information for forecasting is another topic overall. 
%This framework can be understood as an opportunity to apply methodology working with set-valued data and opens up avenues for further research.

\section{Competing interests}
No competing interest is declared.

\section{Author contributions statement}

The core ideas of the paper were developed together and initially inspired by TA. 
Most parts of the paper were drafted and written together, while DK implemented the methodology and wrote the applied section, and TA contributed the embedding in the theory of random sets. 
The specific applied approaches were suggested by DK.

\section{Acknowledgments}
We are most grateful to Civey for the intensive cooperation, the concrete 
implementation of our ideas, sharing of the data and continuous support. 
Financial and general support from the LMU Mentoring
Program (DK) and the Federal Statistical Office of Germany within the cooperation
``Machine Learning in Official Statistics`` (TA and DK) is gratefully acknowledged.
Furthermore, we thank Julian Rodemann and Gunnar K{\"o}nig for the fruitful discussions. %TA danke! Gut, dass Du dran gedacht hast

%%%%%%%%%%%%%%%%%%%%%%%%%%%%%%%%%%%%%%%%%%%%%%%%%%%%%%%%%%%%%%%%%%%%%%%%%%%%%%%%%%%%%%%%%%%%%
%%%%%%%%%%%%%%%%%%%%%%%%%%%%%%%%%%%%%%%%%%%%%%%%

%%%%%%%%%%%%%%%%%%%%%%%%%%%%%
%\bibliographystyle{unsrtnat}
%\bibliographystyle{apastyle}
\bibliographystyle{abbrvnat}
\setcitestyle{authoryear}
\bibliography{references}  %%% Uncomment this line and comment out the ``thebibliography'' section below to use the external .bib file (using bibtex) .

\end{document}